\documentclass[11pt, times,twocolumn]{article}
\usepackage{graphicx}
\usepackage{subfig}
\usepackage{amssymb}
\begin{document}
\topmargin 0pt
\oddsidemargin 0mm
\renewcommand{\thefootnote}{\fnsymbol{footnote}}
\begin{titlepage}

\vspace{5mm}
\begin{center}
{\Large \bf Fermi Velocity Modulation in Graphene by Strain
Engineering} \\

\vspace{6mm} {\large Harihar Behera\footnote{E-mail: harihar@phy.iitb.ac.in;
 harihar@iopb.res.in } and Gautam Mukhopadhyay\footnote{Corresponding author's E-mail: gmukh@phy.iitb.ac.in; g.mukhopa@gmail.com}}    \\
\vspace{5mm}
{\em Department of Physics, Indian Institute of Technology, Powai, Mumbai-400076, India} \\

\end{center}
\vspace{5mm}
\centerline{\bf {Abstract}}
\vspace{5mm}
  Using full-potential density functional theory (DFT) calculations, we found a small asymmetry in the Fermi velocity of electrons and holes in graphene. These Fermi velocity values and their average were found to decrease with increasing in-plane homogeneous biaxial strain; the variation in Fermi velocity is quadratic in strain. The results, which can be verified by Landau level spectroscopy and quantum capacitance measurements of bi-axially strained graphene, promise potential applications in graphene based straintronics and flexible electronics. \\


\end{titlepage}

%%%%***********************************************************************
%\begin{document}
%\begin{titlepage}
%\begin{frontmatter}



%%\begin{abstract}

%\end{abstract}

%%\end{keyword}
%%\end{titlepage}
%%\end{frontmatter}

% \linenumbers

\section{Introduction}
%\label{}
Graphene, a single layer of carbon atoms arranged
in a two-dimensional (2D) hexagonal lattice, displays
remarkable mechanical and electronic properties
promising many applications in nano-devices \cite{1}. The
novel properties of graphene arise from the linear
energy dispersion near the $K$ point of the hexagonal
Brillouin zone (BZ):
\begin{equation}
E - E_F \approx \pm v_F \hbar k
\end{equation}
where $E_F$ is the Fermi energy, $v_F$ is the Fermi
velocity of electrons/holes near the $K$ point of the BZ
and $(\hbar k)$ is the momentum. Many graphene device
characteristics depend on $v_F$. For instance, graphene's
fine structure constant
\begin{equation}
\alpha _G = \frac{e^2}{4\pi \epsilon _0 \hbar v_F}\,\,\,\,\,\,\,\, \mbox{(S. I. units)}
\end{equation}
appears in the graphene device characteristics of
graphene field effect transistor (GFET) \cite{2}; for an
ideal graphene with a uniform channel potential $V_{ch}$,
the quantum capacitance of graphene is given by \cite{3}:
\begin{equation}
C_G =\frac{2e^2k_BT}{\pi (\hbar v_F)^2}\mbox{ln}\left[2\left(1+\cosh\frac{eV_{ch}}{k_BT}\right)\right]
\end{equation}
where $T$ is the temperature and other symbols have
their usual meanings. Thus, any change in $v_F$ would
affect the graphene device characteristics. Although
strain-induced variation in $v_F$ has recently been
observed experimentally \cite{4} by Raman spectroscopy
study of uni-axially strained graphene, the quantitative
dependence of $v_F$ on strain remains unclear. Here, we
report our theoretical investigation on the effect of
biaxial strain on the Fermi velocity of charge carriers
in graphene, which is experimentally known to sustain
in-plane tensile elastic strain in excess of $20\%$ \cite{5}.
Thus, our study mimics the experimental condition
where graphene is supported on an ideal flat
stretchable substrate.


\section{Computational Methods}
We use the DFT based full-potential (linearized)
augmented plane-wave plus local orbital (FP-
(L)APW+lo) method \cite{6} as implemented in the elk
code \cite{7}. For the exchange-correlation term, we use
the Perdew-Zunger variant of local density
approximation (LDA) \cite{8}, the accuracy of which has
been successfully tested in our previous works \cite{9}. For
plane wave expansion in the interstitial region, we
have used $|{\bf G}+{\bf k}|_{max}\times R_{mt} = 9$, where $R_{mt}$
 is the muffintin radius, for deciding the plane wave cut-off. The $k$-point
grid size of $30\times30\times1$ was used for all
calculations. The total energy was converged within
$2\mu$ eV/atom. The 2D hexagonal structure of graphene
was simulated by 3D hexagonal super cell construction
with a large value of c-parameter ($|{\bf c}|= 40$ a.u.). The
application of homogeneous in-plane biaxial $\delta$ strain
was simulated by varying the in-plane lattice parameter
$a (=|{\bf a}| = |{\bf b}|); \delta = (a - a_0)/a_0$, where $a_0$ is the
ground state in-plane lattice constant.

\begin{figure}
 \centering
 %%---start of first subfigure ---
 \subfloat[]{
\label{fig:subfig:a} %% label for first figure
\includegraphics[scale=0.65]{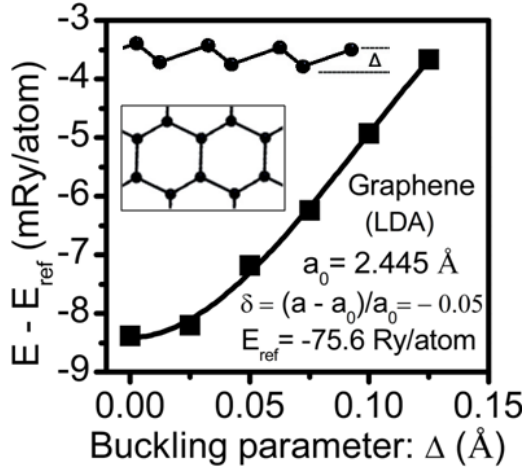}}
 \hspace{0.01in}
 \subfloat[ ]{
 \label{fig:subfig:b} %% label for first figure
 \includegraphics[scale=1.5]{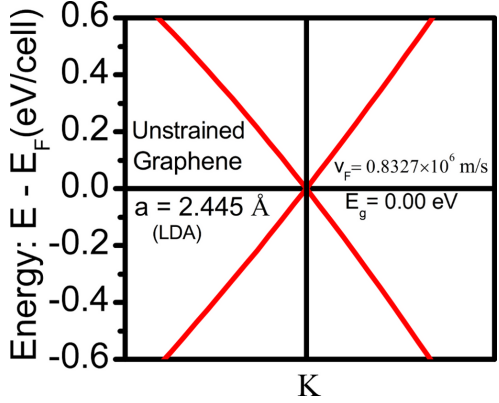}}
\caption{(a) Buckling probe of graphene at $5\%$
compressive biaxial strain using the energy minimization
procedure. Inset shows the side and the top-down views of
buckled graphene in which alternate atoms in a hexagon
reside on two different parallel planes; buckling parameter $\Delta$
is the perpendicular distance between these two planes; for
planar graphene $\Delta = 0.00$\AA. (b) Energy bands of unstrained
graphene near $K$ point of the BZ.
}
\end{figure}
\section{Results and Discussions}
 In Figure 1(a), our calculated results show that
under compressive strain of $5\%$ graphene does not
show any buckling, i.e. its planar structure remains
preserved. Figure 1(b) depicts the linear energy band
dispersion of unstrained graphene near the $K$ point of
the BZ. In Table 1, we compare our calculated values
of $a_0$ and $v_F$ with some reported values.
\begin{table}
\caption{Calculated values of $a_0$ and $v_F$ compared with
reported values.}
\centering
\begin{tabular}{|c|c|p{3.5cm}|}
\hline
 $a_0$ (\AA) & $v_F$ ($10^6$ m/s) &   Remark/Reference \\
\hline
{2.4450} &  0.8327  &  This Work  \\
{2.4595} &  0.833   &  LAPW + GGA \cite{10}  \\
{}       &  0.79    &  Experiment, graphene on graphite substrate \cite{11} \\
{}       &  1.093   &  Experiment, graphene on SiO$_2$/Si substrate \cite{12}   \\
{}       &  0.81    &  Experiment, single walled CNT \cite{13} \\
\hline
\end{tabular}
\end{table}
 Our calculated variation of Fermi velocity of
electrons $v_F(e)$, holes $v_F(h)$ and their average value
$ v_F = \left[v_F(e)+v_F(h)\right]/2$ at different strain values
 are shown in Figure 2. Since $v_F(h)$ values are slightly greater
than $v_F(e)$ values, the band structure of electrons and holes
are not exactly symmetric in qualitative agreement
with a recent experiment \cite{12} and the origin of this
asymmetry is not well understood \cite{12}. The decrease
in Fermi velocity with increasing strain is due to a
reduction in the $\pi$-orbital overlap \cite{4}. Our calculated
data in Figure 2 best fit with the following equations:
\begin{equation}
v_F(e) = 0.82386 - 1.26826\times \delta + 0.71355\times \delta ^2
\end{equation}
\begin{equation}
v_F(h) = 0.84161 - 1.3269\times\delta + 0.76501\times\delta ^2
\end{equation}
\begin{equation}
v_F = 0.83273 - 1.29758\times \delta + 0.73928 \times \delta ^2
\end{equation}
A possible precise measurement of the variation of
$v_F$ with $\delta$ may be carried out with Landau level
spectroscopy study of a bi-axially strained graphene,
since the Landau level spectrum is given by \cite{12}:
\begin{equation}
E_n = E_D + \mbox{sgn(n)}v_F\sqrt{2e\hbar B|n|}
\end{equation}
where $n$ is an integer, $E_D$ is the energy at the Dirac
point, $B$ is the magnetic field perpendicular to
graphene's plane and other symbols have their usual
meanings. Measurement of the quantum capacitance
\cite{3} of bi-axially strained graphene can also be useful.

\begin{figure}
 \centering
 \includegraphics[scale=0.85]{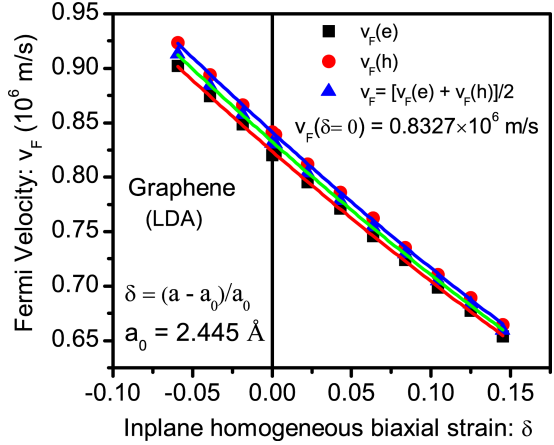}
 \caption{Fermi velocity variation with strain $\delta$.}
\end{figure}

\section{Conclusions}
 There exists a small asymmetry in the Fermi
velocity of electrons and holes in graphene whose
reason is not yet clear. The Fermi velocity in graphene
can be modulated by strain engineering for potential
applications in graphene based flexible electronics.

%% References


\begin{thebibliography}{00}
\bibitem{1} A. H. Castro Neto, et al., Rev. Mod. Phys. {\bf 81}, 109 (2009).
\bibitem{2} F. Schwierz, Nature Nanotechnology {\bf 5}, 487 (2010).
\bibitem{3} H. Xu, et al., Appl. Phys. Lett. {\bf 98}, 133122 (2011).
\bibitem{4} M. Huang, et al., Nano Lett. {\bf 10}, 4074 (2010).
\bibitem{5} C. Lee, et al., Science {\bf 321} (5887), 385 (2008).
\bibitem{6} E. Sj\"{o}stedt, L. Nordstr\"{o}m, D. J. Singh, Solid
State Commun. {\bf 114}, 15 (2000).
\bibitem{7} Freely available at:\\ http://elk.sourceforge.net/
\bibitem{8} P. Perdew, A. Zunger, Phys. Rev. B {\bf 23}, 5048 (1981).
\bibitem{9} H. Behera , G. Mukhopadhyay, J. Phys. Chem. Solids {\bf 73}, 818 (2012); and other works cited in it.
\bibitem{10} M. Gmitra, et al., Phys. Rev. B {\bf 80}, 235431 (2009).
\bibitem{11} G. Li, et al., Phys. Rev. Lett. {\bf 102}, 176804 (2009).
\bibitem{12} K.-C. Chuang, et al., Phil. Trans. R. Soc. A {\bf 366}, 237
(2008).
\bibitem{13} W. Liang, et al., Nature {\bf 411}, 665 (2001).
\end{thebibliography}
%\section*{References}
%\renewcommand\bibname{References}
\bibliographystyle{elsarticle-num}

\end{document}